\documentclass[12pt]{amsart}
\usepackage{amsmath,amssymb}
\usepackage{amsmath}
\usepackage{amsopn}
\usepackage{fixltx2e}
\usepackage{url}

\numberwithin{equation}{section} 
\numberwithin{figure}{section} 

\theoremstyle{plain}
\newtheorem{thm}{Theorem}
\newtheorem{lem}[thm]{Lemma} 
\newtheorem{prop}[thm]{Proposition} 

\theoremstyle{definition}

\theoremstyle{remark}
 \newtheorem*{rem*}{Remark}

\newcommand{\ip}[1]{\langle #1 \rangle}
\newcommand{\Ev}[1]{\E \left( #1 \right)}  
\newcommand{\norm}[1]{\left\Vert#1\right\Vert}
\newcommand{\abs}[1]{\left\vert#1\right\vert}
\newcommand{\set}[1]{\left\{#1\right\}}

\renewcommand{\vec}[1]{\mathbf{#1}}

\newcommand{\bb}[1]{\mathbb{#1}}
\newcommand{\cu}[1]{\mathcal{#1}}

\newcommand{\wh}[1]{\widehat{#1}}

\def\num{\operatorname{Num}}

\def\e{\mathrm e}

\def\im{\mathrm i}
\def\Im{\mathrm{Im}}
\def\half {\frac{1}{2}}
\def\1{{\mathsf 1}}
\def\di{\mathrm d}
\def\grad{\nabla}
\def\wt{\widetilde}

\makeatletter
\def\rightharpoondownfill@{%
    \arrowfill@\relbar\relbar\rightharpoondown}
\def\rightharpoonupfill@{%
    \arrowfill@\relbar\relbar\rightharpoonup}
\def\leftharpoondownfill@{%
    \arrowfill@\leftharpoondown\relbar\relbar}
\def\leftharpoonupfill@{%
    \arrowfill@\leftharpoonup\relbar\relbar}
\newcommand{\xrightharpoondown}[2][]{%
    \ext@arrow 0359\rightharpoondownfill@{#1}{#2}}
\newcommand{\xrightharpoonup}[2][]{%
    \ext@arrow 0359\rightharpoonupfill@{#1}{#2}}
\newcommand{\xleftharpoondown}[2][]{%
    \ext@arrow 3095\leftharpoondownfill@{#1}{#2}}
\newcommand{\xleftharpoonup}[2][]{%
    \ext@arrow 3095\leftharpoonupfill@{#1}{#2}}
\newcommand{\xleftrightharpoons}[2][]{\mathrel{%
    \raise.22ex\hbox{%
        $\ext@arrow 3095\leftharpoonupfill@{\phantom{#1}}{#2}$}%
    \setbox0=\hbox{%
        $\ext@arrow 0359\rightharpoondownfill@{#1}{\phantom{#2}}$}%
    \kern-\wd0 \lower.22ex\box0}%
}
\newcommand{\xrightleftharpoons}[2][]{\mathrel{%
    \raise.22ex\hbox{%
        $\ext@arrow 3095\rightharpoonupfill@{\phantom{#1}}{#2}$}%
    \setbox0=\hbox{%
        $\ext@arrow 0359\leftharpoondownfill@{#1}{\phantom{#2}}$}%
    \kern-\wd0 \lower.22ex\box0}%
} \makeatother

\def\Z{\mathbb Z}

\def\R{\mathbb R}
\def\C{\mathbb C}
\def\E{\mathbb E}

\def\ra{\rightarrow}

\def\ran{\operatorname{ran}}

\def\dist{\operatorname{dist}}   
\def\tr{\operatorname{tr}}    

\def\Re{\operatorname{Re}}
\def\Im{\operatorname{Im}}
\def\tem{\textemdash \ }

\title{Diffusion of wave packets in a Markov random potential}
\author{Yang Kang}
\author{Jeffrey Schenker}
\address{Michigan State University \\ East Lansing, MI 48823}
\email{jeffrey@math.msu.edu}

\dedicatory{Dedicated to J\"urg Fr\"ohlich and Tom Spencer on the occasions of their 60th birthdays.}
\date{August 26, 2008; revised January 22, 2009}
\begin{document}
\maketitle
\pagestyle{plain}
\bibliographystyle{amsplain}

\begin{abstract} We consider the evolution of a tight binding wave packet propagating in a time dependent potential.  If the potential evolves according to a stationary Markov process, we show that the square amplitude of the wave packet converges, after diffusive rescaling, to a solution of a heat equation.
\end{abstract}

\section{Introduction}  It is generally expected that over long times the amplitude of a wave propagating in a weakly disordered background will be well described by a diffusion, at least in dimension $d \ge 3$.  This expectation stems from a picture of wave propagation as a multiple scattering process. Scattering off the disordered background results in random phases, and the build up of these phases over multiple scattering events leads eventually to a loss of coherence in the wave.  Decoherent propagation of the wave may be understood as a classical superposition of reflections from random obstacles.  As long as recurrence does not dominate the evolution, the central limit theorem suggests that the amplitude is given in the long run by a diffusion.

So far it has not been possible to turn this heuristic argument into mathematical analysis without restricting the time scale over which the wave evolution is followed as in \cite{Erdos2007:621, Erdos2007:1, Erdos2008:211}.     One major obstacle is a lack of control over recurrence:  the wave packet may return often to regions visited previously, denying us the independence needed to carry out the central limit argument. Indeed, the phenomenon of Anderson localization indicates that under appropriate conditions recurrence can dominate and cause complete localization of the wave packet.  (It is worth noting that, since random walks are highly recurrent in dimensions $d=1$ and $2$, the above heuristic analysis does not support diffusion in $d=1$ or $2$, dimensions in which localization is proved ($d=1$) and expected ($d=2$) to dominate at any disorder strength.)

A natural way to avoid recurrence difficulties is to bring a time dependence into the disordered background \tem \ we suppose that the environment evolves as the packet propagates.  Here we consider a stochastic environment evolving independently of the wave packet.  A natural assumption in this context is that the background changes in time according to a stationary Markov process.
Such evolution equations have been proposed as an effective model for the propagation of wave packets in optical fibers \cite{Mitra:2001fy}.

We consider here the simplest example of such a wave equation, namely the tight binding Markov-Schr\"odinger equation 
\begin{equation}\label{MAM}
\begin{cases} \im \partial_{t} \psi_{t}(x) \ = \ T \psi_{t}(x) + \lambda v_{x}(\omega(t)) \psi_{t}(x), \\
\psi_{0} \in \ell^{2}(\Z^{d}).
\end{cases}
\end{equation} 
where $T$ is a translation invariant hopping operator on $\ell^{2}(\Z^{d})$, $\omega(t)$ is a Markov process on a measure space $\Omega$ and $v_{x}: \Omega \ra \R$ are measurable functions on $\Omega$.    An elementary, but important,  observation is that,  so long as the time dependent generator $H_{\omega(t)}= T + \lambda v_{x}(\omega (t))$ is uniformly bounded in time,  the non-autonomous problem \eqref{MAM} has a unique solution $\psi_{t}$ for any initial condition $\psi_{0} \in \ell^{2}(\Z^{d})$, given for instance by the norm convergent series
\begin{equation}
\psi_{t} \ = \ \psi_{0} + \sum_{n=1}^{\infty} (-\im )^{n} \int_{\Gamma_{n}(t)}   H_{\omega( r_{1})} \cdots H_{\omega( r_{n})} \psi_{0} \di r_{1} \cdots \di r_{n}, \end{equation}
with $\Gamma_{n}(t) \ = \ \{ (r_{1}, \cdots, r_{n}) \ : \ 0 \le r_{n} \le \cdots \le r_{1} \le t\}$.  The evolution is easily seen to be unitary, $\norm{\psi_{t}} = \norm{\psi_{0}}$.  Thus a sufficient condition for solutions to \eqref{MAM} to exist is that $\norm{T} < \infty$ and $\sup_{x}\sup_{\omega}|v_{x}(\omega)| <\infty$.

We examine diffusion of the wave packet by considering the mean square amplitude $\Ev{|\psi_{t}(x)|^{2}},$ where $\Ev{\cdot}$ denotes averaging with respect to the random paths of the Markov process and also an initial distribution for $\omega(0)$.    Diffusion is characterized by changes in position that scale as the square root of the elapsed time.  Thus it is natural to look at the mean square amplitude in the scaling $t \mapsto \tau t$, $x \mapsto \sqrt{\tau} x$ for a large parameter $\tau$.  Since $x$ is a discrete variable, to accomplish this rescaling we need to convolve $\Ev{|\psi_{t}(x)|^{2}}$ with a function on $\R^{d}$.  That is, we look at \begin{equation}
A_{t}(x) \ = \ \sum_{\xi \in \Z^{d}} u(x-\xi) \Ev{|\psi_{t}(\xi)|^{2}}
\end{equation}
with $u$ a fixed  ``bump function'' centered at $0$. Let us suppose $u \ge 0$ and $\int u\di x=1$, so that $A_{t}(x) \ge 0$ and $\int A_{t}(x) \di x \ = \ \norm{\psi_{0}}^{2}$.  We interpret diffusion for the mean square amplitude as weak convergence of $A_{t}(x)$ under diffusive scaling to a solution of a heat equation.  That is, for suitable test functions $\phi$,
\begin{equation}\label{eq:diffusion!}
\int_{\R^{d}} \phi(x) \tau^{\frac{d}{2}}A_{\tau t} (\sqrt{\tau} x) \di x \ \xrightarrow{\tau \ra \infty} \ \norm{\psi_{0}}^{2} \int_{\R^{d}} \phi(x) \frac{1}{(\pi Dt)^{\frac{d}{2}}}\e^{-\frac{|x|^{2}}{Dt} } \di x
\end{equation}
with $D > 0$.  A sufficient condition for \eqref{eq:diffusion!}, which requires no choice of a bump function, is obtained by a Fourier transform: 
\begin{equation}\label{eq:diffusion!FT}
\sum_{\xi \in \Z^{d}} \e^{- \im \frac{1}{\sqrt{\tau}} \vec{k} \cdot \xi} \Ev{|\psi_{\tau t}(\xi)|^{2}} \ \xrightarrow{\tau \ra \infty} \ \norm{\psi_{0}}^{2} \e^{- t D |\vec{k}|^{2}}, \quad \forall \vec{k} \in \R^{d}.
\end{equation}

Following is a brief history of related studies. Ovchinnikov and Erikman obtained diffusion for a Gaussian Markov (``white noise'') potential \cite{Ovchinnikov:1974eu}.  Pillet obtained results on transience of the wave in related models and derived a Feynman-Kac representation which we use below \cite{Pillet:1985oq}.  The evolution \eqref{MAM} was considered by Tchermentchansev \cite{Tcheremchantsev:1997kl, Tcheremchantsev:1998qe},  who used Pillet's Feynman-Kac formula to show that position moments such as $\sum_{x}|x|^{p} \Ev{ \abs{\psi_{t}(x)}^{2}}$ exhibit diffusive scaling, up to logarithmic corrections.  More precisely, he obtained upper and lower bounds of the form
\begin{equation}\label{eq:scaling}
t^{\frac{p}{2}} \frac{1}{(\ln t)^{\nu_{-}}}\ \lesssim \ \sum_{x}|x|^{p} \Ev{ \abs{\psi_{t}(x)}^{2}} \ \lesssim \  t^{\frac{p}{2}} ( \ln t )^{\nu_{+}} , \quad t \ra \infty.
\end{equation}
In the present paper we obtain \emph{diffusion} \eqref{eq:diffusion!FT} and show that \eqref{eq:scaling} holds for $p=2$ with $\nu_{+}=\nu_{-}=0$.  While completing this manuscript, we learned of recent work of De Roeck, Fr\"ohlich and Pizzo on diffusion for a quantum particle weakly coupled to an array of independent free quantum fields \cite{W.-De-Roeck:rz}.

In the next section we state technical conditions which allow us to derive \eqref{eq:diffusion!FT}.  These conditions are quite general and cover a large number of models of the form \eqref{MAM}.  However, it may be useful to have at least one example in mind.  So, to guide the reader, we close this introduction with a simple example of a potential for which we can derive diffusion.  We call this the ``flip process.''  The state space $\Omega$ of the Markov potential is just $\{-1,1\}^{\Z^{d}}$, and $v_{x}(\omega) = $ evaluation of the $x^{\mathrm{th}}$ coordinate.  So at any time $t$, the potential $v_{x}(\omega(t))$ takes only the values $\pm1$.   Now suppose the process $\omega(t)$ is obtained by allowing each coordinate $v_{x}(\omega)$ to flip sign at the times $t_{1}(x) \le t_{2}(x) \le \cdot$ of a Poisson process, with independent, identical Poisson processes at each site $x$.   For this potential, our result implies diffusion \eqref{eq:diffusion!FT} of the wave amplitude.

\section{Statement of the main result: diffusion of the amplitude}\label{sec:main}
\subsection{Assumptions} We make the following assumptions:
\begin{enumerate}
\item[(1)] (\emph{Existence of the Markov process and invariant measure}): We are given a topological space $\Omega$, a Borel probability measure $\mu$, and a collection $\{\bb{P}_{\alpha} \ : \ \alpha \in \Omega\}$ of probability measures on the path space $\cu{P} =\Omega^{[0,\infty)}$, taken with the $\sigma$-algebra generated by Borel-cylinder sets, such that
\begin{enumerate}
\item (\emph{Paths are right continuous and start at $\alpha$}): For each $\alpha \in \Omega$, with $\bb{P}_{\alpha}$ probability one,  every path $\omega(\cdot)$ is right continuous  and  satisfies $\omega(0) = \alpha$.
\item (\emph{The Markov property holds}):  For any measurable $\cu{A} \subset \cu{P}$ we have
$$\int_{\cu{P}} \bb{P}_{\omega(t)}(\cu{A}) \di \bb{P}_{\alpha}(\omega(\cdot) ) = \bb{P}_{\alpha}(\cu{S}_{t}^{-1}(\cu{A}))$$ where $\cu{S}_{t}$ is the backward shift on $\cu{P}$,  $\cu{S}_{t}\omega(\cdot) = \omega(\cdot +t),$ so
$$\cu{S}_{t}^{-1}(\cu{A}) \ = \ \{ \omega(\cdot) \ : \ \omega(\cdot + t) \in \cu{A} \}.$$
\item (\emph{Invariance of $\mu$}): For any measurable $A \subset \Omega$,
$$\int_{\Omega} \bb{P}_{\alpha}(\omega(t) \in A) \di \mu(\alpha)  \ = \ \mu(A)$$
\end{enumerate}
\end{enumerate}

Let $\E_{\alpha}(\cdot)$ denote expectation with respect to $\bb{P}_{\alpha}$,
\begin{equation}
\E_{\alpha}\left ( F(\omega(\cdot)) \right) \ = \ \int_{\cu{P}}F(\omega(\cdot)) \di \bb{P}_{\alpha}(\omega(\cdot)),
\end{equation}
and similarly
\begin{equation}
\Ev{ \cdot } \ = \ \int_{\Omega} \E_{\alpha}(\cdot) \di \mu(\alpha),
\end{equation}
which is expectation with respect to the probability measure
$\bb{P}(\cu{A})  =  \int_{\Omega} \bb{P}_{\alpha}(\cu{A}) \di \mu(\alpha)$ on $\cu{P}$. By the invariance of $\mu$ under the Markov process we have
\begin{equation}\label{eq:invariantexp}
\Ev{f(\omega(t))} = \int_{\Omega}f(\alpha) \di \mu(\alpha)
\end{equation}
for any $t$ and any $f \in L^{1}(\Omega)$.   

The Markov property, invariance of $\mu$, and right continuity of paths show that
\begin{equation}
S_{t}f(\alpha) \ = \ \E_{\alpha} \left ( f(\omega(t)) \right )
\end{equation}
defines a strongly continuous contraction semi-group on $L^{2}(\Omega)$ (also on $L^{p}(\Omega)$ for any $1\le p\le \infty$).   Indeed, the Markov property clearly shows this is a semi-group, and from the definition we have
\begin{equation}\label{eq:sgip}
\ip{f, S_{t}g}_{L^{2}(\Omega)} \ = \ \Ev{f(\omega(0)) g(\omega(t))},
\end{equation}
so by Cauchy Schwartz and \eqref{eq:invariantexp}
\begin{equation}
|\ip{f, S_{t}g}_{L^{2}(\Omega)}| \ \le \ \norm{f}_{L^{2}(\Omega)} \norm{g}_{L^{2}(\Omega)} 
\end{equation}
from which it follows that $\norm{S_{t}}_{L^{2}\ra L^{2}} \le 1$.   The right continuity of the paths under $\bb{P}_{\alpha}$ now shows that $S_{t}$ is strongly continuous, since  
any $f \in L^{2}(\Omega)$ may be approximated by bounded continuous functions and for bounded continuous $f  : \Omega \ra \R$ we have
\begin{equation}
\norm{ S_{t} f - f}_{L^{2}(\Omega)}^{2} \ = \ \Ev{\abs{f(\omega(t)) - f(\omega(0))}^{2}} \ \ra \ 0
\end{equation}
by dominated convergence.

The adjoint $S_{t}^{\dagger}$ of $S_{t}$  is also a strongly continuous contraction semi-group, given formally  by 
\begin{equation}
S_{t}^{\dagger}f(\alpha) \  = \ \E( f(\omega(0)) | \omega(t) =\alpha),
\end{equation}
where the r.h.s.\ is a conditional expectation.  Of particular importance to us is the generator $B$ of $S_{t}^{\dagger}$  defined by
\begin{equation}
B \psi \ = \ \lim_{t \ra 0^{+}}\frac{1}{t} \left (  \psi - S_{t}^{\dagger} \psi \right )
\end{equation}
on the domain $\cu{D}(B)$ consisting of $\psi$ such that the limit on the r.h.s.\ converges in $L^{2}$ norm. The generator $B$ is \emph{maximally dissipative}, meaning $\Re \ip{\psi, B \psi} \ge 0$ for $\psi \in \cu{D}(B)$ and no extension of $B$ has this property.  It follows that the spectrum of $B$ is contained in the closed right half plane $\{ z \ :   \ \Re z \ge 0\}$.  The adjoint $B^{\dagger}$ of $B$ is the generator for $S_{t}$. 

Note that both $S_{t}$ and $S_{t}^{\dagger}$ satisfy
\begin{equation}
S_{t} 1 \ = \ S_{t}^{\dagger}1 = 1,
\end{equation}
where $1$ denotes the function equal to one everywhere on $\Omega$.
It follows that $1 \in \cu{D}(B)$ and $1 \in \cu{D}(B^{\dagger})$ and that
\begin{equation}\label{eq:eigenvector}
B 1 = B^{\dagger} 1 = 0.
\end{equation}
The orthogonal complement of $1$ is the space of mean zero functions,
\begin{equation}\label{eq:L20}
L^{2}_{0}(\Omega) \ = \ \set{ f \in L^{2}(\Omega) \ : \ \int_{\Omega}f(\alpha) \di \mu(\alpha) =0}.
\end{equation}
From \eqref{eq:eigenvector} it follows that $L^{2}_{0}(\Omega)$ is an invariant subspace for $B$ and $B^{\dagger}$. We require that $B$ is \emph{strictly dissipative} on this space:
 \begin{enumerate}
\item[(2)](\emph{Gap condition and sectoriality of  the generator}): There is $T >0$ such that if $f \in \cu{D}(B)$  and $\int_{\Omega} f(\alpha) \di \mu(\alpha) = 0$, then 
$$
\Re\ip{f, B f}_{L^{2}(\Omega)} \ge \frac{1}{T} \int_{\Omega} \abs{f(\alpha)}^{2} \di \mu(\alpha).
$$
In addition, we require that $B$ is sectorial, namely there is $\gamma < \infty$ such that
$$ \abs{\Im \ip{f, B f}_{L^{2}(\Omega)}} \ \le \ \gamma \Re \ip{f, B f}_{L^{2}(\Omega)} $$
for all $f \in \cu{D}(B)$.
\end{enumerate}
One consequence of the sectorial condition on the generator is that a precise meaning can be given to the formal relation $S_{t}^{\dagger}=\e^{-t B}$ using the Riesz functional calculus \tem \ see \cite[Chapter II, Section 4]{Engel:2000nx}. 

Finally, we require translation invariance for the hopping operator $T$, the Markov process, and the potential $v_{x}(\alpha)$:
\begin{enumerate}
\item[(3)]  (\emph{Translation invariance of the hopping terms}): $T$ is a translation invariant hopping operator on $\ell^{2}(\Z^{d})$, 
$$T \psi(x) \ = \ \sum_{y} h(x-y) \psi(y),
$$ with $h$  a function such that 
\begin{enumerate}
\item (\emph{Self adjointness of $T$}) For every $x$, $h(-x) = h(x)^{*}$.
\item (\emph{Non-degeneracy of $T$}) For each non-zero vector $\vec{k} \in \R^{d}$, there is some $x \in \Z^{d}$ such that $h(x) \neq 0$ and $\vec{k} \cdot x  \ne 0$.
\item (\emph{Smoothness of the symbol}) $ \sum_{x} |x|^{2} \abs{h(x)} \ < \ \infty.$
\end{enumerate}
It follows from (c) that $\widehat h(\vec{k}) = \sum_{x} \e^{-\im \vec{k} \cdot x} h(x)$ is a $C^{2}$ function on the torus $\bb{T}^{d}$. In particular, $T$ is a bounded operator with $\norm{T}_{\ell^{2}(\Z^{d}) \ra \ell^{2}(\Z^{d})} = \max_{\vec{k}} |\widehat h(\vec{k})|.$ 

\item[(4)] (\emph{Translation invariance of the process and invariant measure}): There are $\mu$-measure preserving maps $\sigma_{x}:\Omega \ra \Omega$, $x\in \Z^{d}$, such that
$$ \sigma_{x} \circ \sigma_{y} = \sigma_{x+y} ,$$
and
$$ \bb{P}_{\sigma_{x}(\alpha)}(\cu{T}_{x}(\cu{A})) \ = \ \bb{P}_{\alpha}(\cu{A}),$$
where $\cu{T}_{x}: \cu{P} \ra \cu{P}$ is the map $\cu{T}_{x}(\omega)(\cdot) = \sigma_{x}(\omega(\cdot)).$
\item[(5)] (\emph{Translation covariance, boundedness and non-degeneracy of the potential}): The functions $v_{x} :\Omega \ra \R$ are bounded, translation covariant
$$ \quad v_{x} = v_{0} \circ \sigma_{x},$$
mean zero
$$ \int_{\Omega} v_{x}(\alpha) \di \mu(\alpha) = 0,$$
and there is $\chi > 0$ such that
for all $x, y \in \Z^{d}$, $x \neq y$,
\begin{equation}\label{eq:nondeg}
\norm{B^{-1} (v_{x}- v_{y})}_{L^{2}(\Omega)} \ge \chi.
\end{equation}
\end{enumerate}
Since the Markov process is translation invariant, $B$ commutes with the translations $T_{x}f(\alpha) = f(\sigma_{x}\alpha)$ of $L^{2}(\Omega)$.  Thus \eqref{eq:nondeg} is equivalent to
\begin{equation}
\norm{B^{-1}(v_{x}- v_{0})}_{L^{2}(\Omega)} \ge \chi.
\end{equation}
for all $x \in \Z^{d}$, $x \neq 0$.

The condition $\int v_{x} \di\mu =0$ can always be obtained by putting the mean of $v_{x}$ into the diagonal part of the hopping term.  Likewise, by absorbing the normalization into the disorder strength $\lambda$, we may assume that $\norm{v_{x}}_{L^{2}(\Omega)}=1$.

A very general class of models, which includes the flip model described above, is obtained by taking $\Omega = S^{\Z^{d}}$ for some set $S \subset \R$ and supposing that \emph{each coordinate} $\omega(x)$ of $\omega \in \Omega$ undergoes an \emph{independent Markov process}, with the processes at different sites \emph{identically distributed}.   We then set $v_{x}(\omega)=\omega(x)$.  In this case, the generator $B$ is the sum of the individual generators for the processes at each site \tem \ more precisely the Friedrichs extension of that sum defined on the domain of functions depending on only finitely many coordinates.  The above conditions are easily translated into conditions on the individual generator of the Markov process on $S$  for each coordinate $\omega(x)$.    For these models, the condition \eqref{eq:nondeg} is \emph{trivial} since, by the independence of different coordinates and translation invariance, we have 
\begin{equation}
\norm{B^{-1} (v_{x}- v_{y})}^{2}_{L^{2}(\Omega)} = \norm{B^{-1} v_{x}}^{2}_{L^{2}(\Omega)}+ \norm{B^{-1} v_{y}}^{2}_{{L^{2}(\Omega)}} = 2 \norm{B^{-1} v_{0}}^{2}_{{L^{2}(\Omega)}} , \quad x \neq y.
\end{equation}

\subsection{Main result}The wave function $\psi$ satisfies a \emph{linear} equation, but the square amplitude $\abs{\psi}^{2}$ is \emph{quadratic} in $\psi$.  To obtain a linear equation for the evolution of $\abs{\psi}^{2}$, we consider the density matrix
\begin{equation}
\rho_{t}(x,y) \ = \ \psi_{t}(x) \psi_{t}(y)^{*},
\end{equation}
which satisfies the evolution equation 
\begin{equation}\label{MAMDM}
\partial_{t} \rho_{t}(x,y) \ = \ - \im \sum_{\zeta} h(\zeta) \left[ \rho_{t}(x-\zeta,y) - \rho_{t}(x,y+\zeta) \right ]
-\im \lambda \left ( v_{x}(\omega(t)) - v_{y}(\omega(t)) \right ) \rho_{t}(x,y).
\end{equation}
Note that $
\abs{\psi_{t}(x)}^{2}  =  \rho_{t}(x,x).$   

More generally, we may consider the evolution equation \eqref{MAMDM} to be the basic dynamical problem, with an arbitrary initial condition $\rho_{0}(x,y)$.  The natural setting is for $\rho_{0}$ to be a \emph{density matrix}, namely 
\begin{multline}
\rho_{0} \in \cu{DM} \ := \ \left \{ \rho: \Z^{d} \times \Z^{d} \ra \C \ : \ \rho \text{ is the kernel of a non-negative definite,} \right . \\
\left.  \text{trace class operator on $\ell^{2}(\Z^{d})$} \right \}.
\end{multline}
By virtue of the unitarity of the evolution \eqref{MAM}, the space $\cu{DM}$ is preserved by the evolution \eqref{MAMDM}, as is the trace $\tr \rho_{0} = \sum_{x}\rho_{0}(x,x)$.

Under the assumptions outlined in the previous subsection, we have the following
  \begin{thm}\label{thm:main}
Any solution to \eqref{MAMDM} with initial condition $\rho_{0} \in \cu{DM}$ satisfies
\begin{equation}\label{eq:main}
\lim_{\tau \ra \infty }\sum_{x} \e^{-\im \frac{\vec{k}}{\sqrt{\tau}} \cdot x} \Ev{\rho_{\tau t} (x,x)} \ = \  [\tr \rho_{0} ] \e^{-t \sum_{i,j}D_{i,j}(\lambda) \vec{k}_{i} \vec{k}_{j}},
\end{equation}
with $D_{i,j} = D_{i,j}(\lambda)$ a positive definite matrix. Near $\lambda =0$, $D_{i,j}(\lambda)$ has an asymptotic expansion 
\begin{equation}\label{eq:asdiff}
D_{i,j}(\lambda) = \frac{1}{\lambda^{2}} \left ( D_{i,j}^{0} + O(\lambda) \right ) .\end{equation}
If furthermore $ \sum_{x} \abs{x}^{2} \rho_{0}(x,x) < \infty$, then
\begin{equation}\label{eq:x2}
\lim_{t \ra \infty} \frac{1}{t} \sum_{x} \abs{x}^{2} \Ev{\rho_{t}(x,x)} \ = \ [\tr \rho_{0}] \sum_{i}D_{i,i}.
\end{equation}
\end{thm}
\begin{rem*} As the proof will show, \eqref{eq:main} holds also for an initial condition $\rho_{0}$ which is the kernel of a non-positive definite trace class operator. 
\end{rem*}

\section{Augmented space analysis}
\subsection{Augmented space and Pillet's Feynman-Kac formula}

The starting point of our analysis is a ``Feynman-Kac'' formula due to Pillet  \cite{Pillet:1985oq} which expresses  $\Ev{\rho_{t}(x,x)}$ as a matrix element of a contraction semigroup on an augmented Hilbert space,
 \begin{equation}
 \cu{H} \ := \ L^{2}(\Z^{d} \times \Z^{d} \times \Omega),
\end{equation}
where $\Z^{d} \times \Z^{d} \times \Omega$ is taken with the measure $M$
\begin{equation} \int f(x,y,\omega) \di M(x,y,\omega) \ = \ \sum_{x,y} \int_{\Omega} f(x,y,\omega) \di \mu(\omega).
\end{equation}
We think of a vector $\Psi \in \cu{H}$ as a   ``random density matrix,'' at least if $\Psi(\cdot, \cdot, \omega)$ is the kernel of a non-negative definite trace class operator for $\mu$ almost every $\omega$.    We also think of $\cu{H}$ as the tensor products $\ell^{2}(\Z^{d}) \otimes \ell^{2}(\Z^{d}) \otimes L^{2}(\Omega)$ or $\ell^{2}(\Z^{d} \times \Z^{d}) \otimes L^{2}(\Omega)$, using the notations
\begin{equation}
[\psi \otimes \phi \otimes f] (x,y,\omega) = \psi(x) \phi(y) f(\omega) , \quad \psi, \phi \in \ell^{2}(\Z^{d}) , \ f \in L^{2}(\Omega) ,
\end{equation}
and
\begin{equation}
[\rho \otimes f] (x,y,\omega) = \rho(x,y) f(\omega) , \quad \rho \in \ell^{2}(\Z^{d} \times \Z^{d}) , \ f \in L^{2}(\Omega).
\end{equation}

The Feynman-Kac-Pillet formula  basic to our work is 
\begin{equation}
	\E(\rho_t (x,y)) = \ip{ \delta_{x}\otimes \delta_{y} \otimes 1,  e^{-t(\im K + \im \lambda V +B) } \rho_0 \otimes 1}_{\cu{H}}, \label {E(Rho)}
\end{equation}
where
\begin{equation}
K \Psi(x,y,\omega) \ = \    \sum_{\zeta} h(\zeta) \left[ \Psi(x-\zeta,y,\omega) - \Psi(x,y+\zeta,\omega) \right ],
\end{equation}
\begin{equation}
V \Psi(x,y,\omega) \ = \  \left ( v_{x}(\omega) -v_{y}(\omega) \right ) \Psi(x,y,\omega),
\end{equation}
and the Markov generator $B$ acts on $\cu{H}$ as a multiplication operator with respect to the first two coordinates, that is 
\begin{equation}
B [\rho \otimes f] \ = \ \rho \otimes (Bf), \quad \rho \in \ell^{2}(\Z^{d} \times \Z^{d}) , \ f \in L^{2}(\Omega).
\end{equation}
In particular, we have
\begin{equation}
\E(\rho_{t}(x,x)) = \ip{ \delta_{x}\otimes \delta_{x} \otimes 1,  e^{-tL} \rho_{0} \otimes 1}_{\cu{H}}, \label {E(Rho2)}
\end{equation}
where $L = \im K + \im \lambda V + B$.  This equation
relates the mean square amplitude of the time dependent dynamics \eqref{MAMDM} to spectral properties of the non-self adjoint operator $L$.   

\subsection{Fourier Transform} 
To perform a spectral analysis of $L$, it is useful to note that $L$ commutes with a group of translations on $\cu{H}$ \tem \ a fact which encodes the distributional invariance of \eqref{MAM} under translations.   Specifically, if we let $S_{\xi}$ denote a simultaneous shift of position and disorder,
$$S_\xi \Psi(x, y,  \omega)  = \Psi(x -\xi, y - \xi,  \sigma_{\xi}\omega),$$
then we have
\begin{prop}
 $$S_\xi  K =  K S_\xi , \quad S_{\xi} V = V S_{\xi}, \quad \text { and } \quad S_\xi  B = B S_\xi.$$ 
\end{prop}
\begin{proof} The first two identities follow directly from the definitions of $K$ and $V$; the last follows from the assumed translation invariance of the measure $\mu$.
\end{proof}
As $K$, $V$ and $B$ commute with a representation of the additive group $\Z^{d}$, we may simultaneously partially diagonalize them by a Fourier transform.  In the present context a useful transformation is the following unitary map:
\begin{multline}
\wh{ \Psi}(x,\omega, \vec{k})  = \sum_{\xi} \e^{-i \vec{k} \cdot \xi}S_{\xi} \Psi (x, 0, \omega)   = \sum_{\xi} \e^{-i \vec{k} \cdot \xi}  \Psi (x -\xi , -\xi, \sigma_{\xi} \omega), \\
 \wh{\bullet}  :   L^2(\Z^{2d} \times \Omega) \to   L^2(\Z^{d} \times \Omega \times \bb{T}_d)
\end{multline}
where $\bb{T}_d = [0, 2\pi )^{d}$ is the $d$ torus. One may easily compute that
\begin{align}
    \wh{K} \wh{\Psi}(x,\omega,\vec{k}) := \wh{ K  \Psi} (x, \omega , \vec{k}) &=  \sum_{\zeta} h(\zeta) \left [ 
     \wh{\Psi}(x-\zeta,\omega, \vec{k}) - \e^{-\im \vec{k} \cdot \zeta}
     \wh{\Psi}(x-\zeta, \sigma_{\zeta}\omega, \vec{k}) \right ] \ , \\
    \wh{V} \wh{\Psi}(x,\omega, \vec{k}) :=  \wh{V \Psi} (x, \omega, \vec{k})  &=    (v_{x}(\omega) -v_{0}(\omega)) \wh{ \Psi} (x , \omega, \vec{k}) \ , \intertext{and}
     \wh{B \Psi} (x, \omega, \vec{k}) &= B  \wh{\Psi} (x , \omega, \vec{k}),
\end{align}
where $B$ is understood to act as a multiplication operator with respect to $x \in \Z^{d}$ and $\vec{k} \in \bb{T}^{d}$.

The operators $\wh{K}$, $\wh{V}$ and $B$ act fiberwise over the torus $\bb{T}^{d}$ \tem \ i.e., they act as multiplication operators with respect to the coordinate $\vec{k}$.  Thus, eq.\ \eqref{E(Rho)} may be transformed into
\begin{equation}
	\E(\rho_t (x,y)) = \int_{\bb{T}^{d}} \e^{-\im \vec{k} \cdot y} \ip{ \delta_{x-y} \otimes 1, \e^{-t \wh{L}_{\vec{k}}} \wh{\rho}_{0;\vec{k}}\otimes 1}_{L^{2}(\Z^{d} \times \Omega)} \di \ell(\vec{k}),\label {E(Rho)FT}
\end{equation}
where $\ell$ denotes normalized Lebesgue measure on the torus $\bb{T}^{d}$,
\begin{equation}
\wh{\rho}_{0;\vec{k}}(x) \ = \ \sum_{y} \e^{-\im \vec{k}\cdot y} \rho_{0}(x-y,-y) ,
\end{equation}
and
\begin{equation}
\wh{L}_{\vec{k}}  \ := \ \im \wh{K}_{\vec{k}} + \im \lambda \wh{V} + B
\end{equation}
with $\wh{V} \phi(x,\omega) \ = \  (v_{x}(\omega)- v_{0}(\omega)) \phi(x,\omega)$ 
and
\begin{equation}
\wh{K}_{\vec{k}}\phi(x,\omega) =  \sum_{\zeta} h(\zeta) \left [ 
     \phi(x-\zeta,\omega) - \e^{-\im \vec{k} \cdot \zeta}
     \phi(x-\zeta, \sigma_{\zeta}\omega) \right ] .
\end{equation}
In particular,
\begin{equation}
\sum_{x} \e^{-\im \vec{k} \cdot x} \Ev{\rho_{t}(x,x)} \ = \  \ip{ \delta_{0} \otimes 1, \e^{-t \wh{L}_{\vec{k}}} \wh{\rho}_{0;\vec{k}}\otimes 1}_{L^{2}(\Z^{d} \times \Omega)} \label {E(Rho2)FT}.
\end{equation}
This equation is the starting point for our proof of Theorem \ref{thm:main}.  It indicates that the diffusive limit on the l.h.s.\ of \eqref{eq:main} can be studied via a spectral analysis of the semi-group $\e^{-t \wh{L}_{\vec{k}}}$ for $\vec{k}$ in a neighborhood of $0$.

\section{Spectral analysis of $\wh{L}_{\vec{k}}$ and the proof of Theorem \ref{thm:main}}
In this section inner products and norms are taken in the space $L^{2}(\Z^{d} \times \Omega)$ unless otherwise indicated.

\subsection{Spectral analysis of $\wh{L}_{\vec{0}}$}
To begin, let us consider $\wh{L}_{\vec{k}}$ with $\vec{k}=0$.  A preliminary observation is that
\begin{equation}
\wh{L}_{\vec{0}} \delta_{0}\otimes 1 \ = \ 0.
\end{equation}
Ultimately, this identity is a consequence of the fact that $\sum_{x}\Ev{\tr \rho_{t}}$ is constant in time.   

Let $P_{0}$ denote orthogonal projection of $L^{2}(\Z^{d}\times \Omega)$ onto the space $\cu{H}_{0}= \ell^{2}(\Z^{d})\otimes \{1\}$ of ``non-random'' functions,
\begin{equation}
P_{0}\Psi(x) \ = \ \int_{\Omega } \Psi(x,\omega) \di \mu(\omega).
\end{equation}Then $P_{0}^{\perp}=(1-P_{0})$ is the projection onto mean zero functions 
\begin{equation}
\cu{H}^{\perp}_{0} = \set{ \Psi(x,\omega) \ : \ \int_{\Omega} \Psi(x,\omega) \di\mu(\omega) = 0}.
\end{equation} 
The block decomposition of $\wh{L}_{\vec{0}}$ with respect to the
direct sum $\cu{H}_{0} \oplus \cu{H}_{0}^{\perp}$ is of the form
\begin{equation}\label{eq:blockform}
\wh{L}_{\vec{0}}  \ = \ \begin{pmatrix}
0 & \im \lambda P_{0} \wh{V} \\
\im \lambda  \wh{V} P_{0} & \im \wh{K}_{\vec{0}} + B + \im \lambda P_{0}^{\perp} \wh{V} P_{0}^{\perp}
\end{pmatrix}.
\end{equation}
Indeed, it follows from the definition of $\wh{K}_{\vec{0}}$ that
\begin{equation}
\wh{K}_{\vec{0}} P_{0} = P_{0} \wh{K}_{\vec{0}} =0 ,
\end{equation}
and we have seen in \S\ref{sec:main} that
\begin{equation}
P_{0}B= B P_{0} =0.
\end{equation}
(The identity $P_{0} B =0$ follows since $B^{\dagger}P_{0}=0$.) Thus, 
$\ran P_{0}^{\perp} = \cu{H}^{\perp}_{0}$ is an invariant subspace for $\im \wh{K}_{\vec{0}} +B$ and $\im \wh{K}_{\vec{0}} + B$ vanishes on $\ran P_{0} = \cu{H}_{0}$.  Since the potentials $v_{x}(\omega)$ are mean zero, $P_{0} \wh{V} P_{0}=0$ and \eqref{eq:blockform} follows.

Using the block decomposition \eqref{eq:blockform} we now prove the following
\begin{lem}\label{lem:L0}
For each $\lambda >0$ there is $\delta_{\lambda} >0$, with $\delta_{\lambda} \ge c \lambda^{2}$ for $\lambda$ small, such that
\begin{equation}
\sigma(\wh{L}_{\vec{0}}) \ = \ \{0\} \cup \Sigma_{+} 
\end{equation}
where
\begin{enumerate}
\item $0$ is a non-degenerate eigenvalue, and
\item $\Sigma_{+} \subset \set{z \ : \ \Re z >\delta_{\lambda}}.$
\end{enumerate}
\end{lem}
\begin{proof}
First note that $\Re \wh{L}_{\vec{0}} = \Re B \ge 0$ in the sense of quadratic forms. Thus by the sectoriality of $B$ 
\begin{equation}
\abs{\Im{\langle \Phi, \wh{L}_{\vec{0}} \Phi \rangle}} \ \le \ \norm{\wh{K}_{\vec{0}} + \lambda \wh{V}} + \abs{\Im{\langle \Phi, B \Phi \rangle }} \ \le  \
\ \|\wh{h}\|_{\infty} + 2\lambda + \gamma \Re \langle \Phi, \wh{L}_{\vec{0}} \Phi \rangle,
\end{equation}
if $\norm{\Phi}=1$. It follows that the \emph{numerical range} of $\wh{L}_{\vec{0}}$, $\num(\wh{L}_{\vec{0}}) = \set{ \langle \Phi, \wh{L}_{\vec{0}} \Phi \rangle \ : \ \norm{\Phi}=1}$ is contained in
\begin{equation}\label{eq:numrange}
\cu{N}_{+} \ := \ \set{ z \  : \ \Re z \ge 0 \text{ and } |\Im z| \le \|\wh{h}\| _{\infty} +2 \lambda + \gamma \Re z}.
\end{equation}
Since $\sigma(\wh{L}_{\vec{0}}) \subset \num(\wh{L}_{\vec{0}})$, we may restrict our attention to $z \in \cu{N}_{+}$.

Now  fix $z  \in \cu{N}_{+}$ and consider the equation
\begin{equation}\label{eq:resolvequ}
(\wh{L}_{\vec{0}}-z)  \begin{pmatrix} \phi\otimes 1 \\
\Phi \end{pmatrix}  \ = \ \begin{pmatrix}
-z & \im \lambda P_{0} \wh{V} \\
\im \lambda \wh{V} P_{0} & P_{0}^{\perp} \wh{L}_{\vec{0}}P_{0}^{\perp} - z
\end{pmatrix} \begin{pmatrix} \phi\otimes 1 \\
\Phi \end{pmatrix} \ = \ \begin{pmatrix} \psi\otimes 1 \\ \Psi
\end{pmatrix} ,
\end{equation}
for $(\phi\otimes 1, \Phi) \in \cu{H}_{0}\oplus \cu{H}_{0}^{\perp}$ given
$(\psi\otimes 1, \Psi) \in \cu{H}_{0} \oplus \cu{H}_{0}^{\perp}$. 
By the gap condition on $B$,
\begin{equation}
\Re P_{0}^{\perp} \wh{L}_{\vec{0}}P_{0}^{\perp} \ = \ \Re ( \im \wh{K}_{\vec{0}} + B + \im \lambda P_{0}^{\perp} \wh{V} P_{0}^{\perp} ) \ \ge \ \frac{1}{T} P_{0} ^{\perp},
\end{equation}
so the second of the two equations \eqref{eq:resolvequ} may be solved provided $\Re z < \frac{1}{T}$ to give
\begin{equation}
\Phi \ = \ (P_{0}^{\perp} \wh{L}_{\vec{0}}P_{0}^{\perp} -z )^{-1} \left[  \Psi - \im \lambda \wh{V} \phi\otimes 1 \right ].
\end{equation}
Thus the first equation of \eqref{eq:resolvequ} reduces to
\begin{equation}
\left[ \Gamma(z)  - z \right ]  \phi\otimes 1
\ = \ \psi\otimes 1 - P_{0}\wh{V}  (P_{0}^{\perp} \wh{L}_{\vec{0}}P_{0}^{\perp}-z )^{-1} \Psi,
\end{equation}
with
\begin{equation}
\Gamma(z) \ = \ \lambda^{2} P_{0}\wh{V} ( P_{0}^{\perp} \wh{L}_{\vec{0}}P_{0}^{\perp} -z )^{-1}  \wh{V} P_{0}.
\end{equation}

Thus $\wh{L}_{\vec{0}}-z$ is boundedly invertible, for $\Re z < \frac{1}{T}$, if and only if $z \not \in \sigma(\Gamma(z)).$  However, 
\begin{multline}
\Re \ip{\phi \otimes 1, \Gamma(z)\phi \otimes 1}  \\ = \  \lambda^{2}\ip{\phi \otimes 1, \wh{V} (P_{0}^{\perp}\wh{L}_{\vec{0}}^{\dagger}P_{0}^{\perp}  - z^{*})^{-1} \left ( \Re B - \Re z \right ) (P_{0}^{\perp}\wh{L}_{\vec{0}}P_{0}^{\perp} - z)^{-1}  \wh{V} \phi \otimes 1}
\\
\ge \ \lambda^{2}\left ( \frac{1}{T} - \Re z \right )  \norm{ \left ( P_{0}^{\perp}\wh{L}_{\vec{0}}P_{0}^{\perp}  - z \right )^{-1}  \wh{V}  \phi\otimes 1}^{2} \\
= \ \lambda^{2}\left ( \frac{1}{T} - \Re z \right )  \norm{ \left (B^{-1}P_{0}^{\perp} (\wh{L}_{\vec{0}}- z)P_{0}^{\perp} \right )^{-1} B^{-1} \wh{V} \phi\otimes 1}^{2},
\end{multline}
where the inverse of $B$ is well defined since $\wh{V} \phi\otimes 1 \in \cu{H}^{\perp}_{0} = \ran P_{0}^{\perp}$.  Furthermore $B^{-1}$ is bounded on $\cu{H}^{\perp}_{0}$, with $\norm{B^{-1}P_{0}^{\perp}} \le T$.  Thus $B^{-1}P_{0}^{\perp} (\wh{L}_{\vec{0}}- z)P_{0}^{\perp}$ is bounded, 
\begin{equation}
\norm{B^{-1}P_{0}^{\perp} (\wh{L}_{\vec{0}}- z)P_{0}^{\perp}} \le 1+ T \left ( \|\wh{h}\|_{\infty} + 2 \lambda + |z| \right ),
\end{equation}
and it follows that
\begin{multline}
\Re \ip{\phi \otimes 1 , \Gamma(z)\phi \otimes 1}_{\ell^{2}(\Z^{d})}  \ge \lambda^{2} \left ( \frac{1}{T} - \Re z \right ) \frac{1}{\left [ 1 +T \left ( \|\wh{h}\|_{\infty} + 2 \lambda + |z| \right ) \right ]^{2} } 
 \norm{ B^{-1} \wh{V}  \phi \otimes 1}^{2}
 \\ \ge \lambda^{2}\chi^{2}\left ( \frac{1}{T} - \Re z \right ) \frac{1}{ \left [ 1 +T \left ( \|\wh{h}\|_{\infty} + 2 \lambda + |z| \right ) \right ]^{2}} 
\sum_{x \neq 0} |\phi(x)|^{2},
\end{multline}
where we have made use of the non-degeneracy assumption on the potential.

For $z \in \cu{N}_{+}\cap \set{ \Re z < \frac{1}{T}}$ we have
\begin{equation}
 \|\wh{h}\|_{\infty} + 2 \lambda + |z| \le 2 \|\wh{h}\|_{\infty} + 4 \lambda + (1 + \gamma) \frac{1}{T}.
\end{equation}
Thus
\begin{equation}
\Re \Gamma(z) \ge \frac{1 - T \Re z}{T}  \left [ \frac{\lambda \chi}{ 2 + \gamma +  2 T \|\wh{h}\|_{\infty} + 4 T \lambda  } \right ]^{2}  (1 - \Pi_{0}),
\end{equation}
with $\Pi_{0}$ the projection of $\cu{H}_{0}$ onto $\delta_{0}$.  It follows that 
\begin{enumerate}
\item $\Gamma(0)$ is invertible off the range of $\Pi_{0}$, so $0$ is a non-degenerate eigenvalue of $\wh{L}_{\vec{0}}$,
\item $\Gamma(z) - z$ is invertible if $z \in \cu{N}_{+}$, $z \neq 0$,
\begin{equation}\label{eq:maxmin}
\Re z \ \le \   \frac{r}{T} \ , \quad \text{and} \quad \Re z \le  \frac{1-r}{T}  \left [ \frac{\lambda \chi}{ 2 + \gamma +  2 T \|\wh{h}\|_{\infty} + 4 T \lambda } \right ]^{2},
\end{equation}
for some $r \in (0,1)$.
\end{enumerate}
Optimizing the choice of $r$ gives the following explicit expression for $\delta_{\lambda}$:
\begin{equation}
\delta_{\lambda} \ = \  \frac{1}{T} \frac{\lambda^{2} \chi^{2}} {\left ( 2 + \gamma +  2 T \|\wh{h}\|_{\infty} + 4 T \lambda\right )^{2} + \lambda^{2} \chi^{2}}. \qedhere
\end{equation}
  \end{proof}
  
  The spectral gap $\delta_{\lambda}$ has consequences for the dynamics of the semi-group:
\begin{lem} \label{lem:L0dynamics} Let $Q_{0}= $ orthogonal projection onto $\delta_{0} \otimes 1$  in $L^{2}(\Z^{d}\times \Omega)$.  Then $\e^{-t \wh{L}_{\vec{0}}} (1-Q_{0})$
is a contraction semi-group on $\ran (1-Q_{0})$, and for all sufficiently small $\epsilon >0$  there is $C_{\epsilon} > 0$ such that
\begin{equation}\label{eq:expdecay}
\norm{\e^{-t \wh{L}_{\vec{0}}} (1- Q_{0})} \ \le \ C_{\epsilon} \e^{-t (\delta_{\lambda} -\epsilon)}
\end{equation}
\end{lem}
\begin{proof}That $\e^{-t \wh{L}_{\vec{0}}} (1-Q_{0})$ is a semi-group on $\ran (1-Q_{0})$ with generator $\wh{L}_{\vec{0}}^{(0)}= \wh{L}_{\vec{0}} |_{\ran (1- Q_{0})}$ is clear.
Since $\sigma(\wh{L}_{\vec{0}}) \subset \{ \Re z \ge \delta_{\lambda}\}$, it is known that a sufficient condition for \eqref{eq:expdecay} to hold is for $\e^{-t \wh{L}_{\vec{0}}} (1-Q_{0})$  to be an analytic semi-group (see \cite[Chapter IV Corollary 3.12]{Engel:2000nx}).   A necessary and sufficient condition for $\e^{-t \wh{L}_{\vec{0}}} (1-Q_{0})$ to be analytic, given that $\wh{L}_{\vec{0}}$ is maximally dissipative, is that the following estimate
\begin{equation}\label{eq:sectorial2}
\norm{\frac{1}{\zeta - \wh{L}_{\vec{0}}^{(0)}}}_{\ran (1 - Q_{0})} \le \frac{C}{|\Im \zeta|}
\end{equation}
holds
for all $\zeta \in \C$ with $\Re \zeta <0$.  (See \cite[Chapter II Theorem 4.6]{Engel:2000nx}.)

For any invertible closed operator $A$, we have  $\norm{A^{-1}} \le 1/ \inf \set{|z|  : z \in \num(A)}$, with $\num(A)$ the numerical range of $A$.    Since  $\num(\wh{L}_{\vec{0}}^{(0)})\subset \cu{N}_{+}$, with $\cu{N}_{+}$ as in \eqref{eq:numrange}, we have
\begin{equation}
\norm{\frac{1}{\zeta-\wh{L}_{\vec{0}}^{(0)}}}_{\ran(1- Q_{0})} \le \frac{1}{ \dist(\zeta, \cu{N}_{+})}.
\end{equation}
It follows that \eqref{eq:sectorial2} holds for all $\zeta$ outside a rectangle of the form $R=[-a,0] + \im [-M,M]$.  On the other hand, by Lemma \ref{lem:L0},  this rectangle $R$ is  contained in the resolvent set of $\wh{L}_{\vec{0}}^{(0)}$.  Since $R$ is compact, we  have 
\begin{equation} 
\sup_{\zeta \in R} \norm{\frac{1}{\zeta -\wh{L}_{\vec{0}}^{(0)}}}_{\ran(1 - Q_{0})} < \infty ,
\end{equation}
which is stronger than \eqref{eq:sectorial2} for $\zeta \in R$.
\end{proof}

\subsection{Analytic perturbation theory for $\wh{L}_{\vec{k}}$} 
Now that we have established a strict spectral gap for $\wh{L}_{\vec{0}}$, it follows that the gap persists in the spectrum of $\wh{L}_{\vec{k}}$ for $\vec{k}$ sufficiently small.  Indeed,
\begin{equation}
[\wh{L}_{\vec{k}}- \wh{L}_{\vec{0}}]\phi(x,\omega) \ = \  \sum_{\zeta} h(\zeta) (1- \e^{-\im \vec{k} \cdot \zeta}) \phi(x-\zeta,\sigma_{\zeta}\omega),
\end{equation}
so
\begin{equation}
\norm{\wh{L}_{\vec{k}} - \wh{L}_{\vec{0}}} \ \le \ c |\vec{k}|.
\end{equation}
Thus, an immediately corollary of Lemma ~\ref{lem:L0}  is 
\begin{lem}\label{lem:analytic}
If  $|\vec{k}|$ is sufficiently small, the spectrum of $\wh{L}_{\vec{k}}$ consists of:
\begin{enumerate}
\item A non-degenerate eigenvalue $E(\vec k)$ contained in $H_0 = \{ z\, :\, |z| < c |\vec{k}|\}.$ 
\item The rest of the spectrum is contained in the half plane $ H_1 = \{ z : \Re z > \delta_{\lambda} - c |\vec{k}| \}$ such that $H_{0} \cap H_1 = \emptyset$. 
\end{enumerate}
Furthermore, $E(\vec{k})$ is $C^{2}$ in a neighborhood of $0$,
\begin{equation}
E(\vec 0) = 0, \quad \nabla E(\vec 0) = 0, 
\end{equation}
and
\begin{multline}\label{eq:explicit}
\partial_{i} \partial_{j} E(\vec 0) \ = \ 2 \Re \ip{ \partial_{i} \wh{K}_{\vec{0}} \delta_{0} \otimes 1, [\wh{L}_{\vec{0}}]^{-1}  \partial_{j} \wh{K}_{\vec{0}} \delta_{0} \otimes 1} \\
= \  2 \Re \sum_{x,y \in \Z^{d}}x_{i} y_{j} \overline{h(x)} h(y) \ip{ \delta_{x} \otimes 1 ,[\Gamma(0)]^{-1} \delta_{y}\otimes 1} ,
\end{multline}
with
\begin{equation}
\Gamma(0) = \lambda^{2}P_{0}\wh{V}  ( P_{0}^{\perp} \wh{L}_{\vec{0}}P_{0}^{\perp}  )^{-1} \wh{V} P_{0}
\end{equation}
as in the proof of Lemma \ref{lem:L0}.  In particular, $\partial_{i} \partial_{j} E(\vec 0)$ is positive definite.  
\end{lem}
\begin{proof}
These are all standard facts from analytic perturbation theory \tem \ see for instance \cite{Kato:1995sf}. A sketch of the proof is as follows.
First, it is a general fact that the spectrum moves no further than the norm of the perturbation, so
\begin{equation}
\sigma(\wh{L}_{\vec{k}}) \subset \set{z \ : \ |z| < c |\vec{k}|} \cup \set{ z \ : \ \Re z > \delta_{\lambda}- c |\vec{k}|}.
\end{equation}
For sufficiently small $\vec{k}$ the two sets on the r.h.s.\ are disjoint and we may fit a contour $\cu{C}$ which winds around the origin between them.  The (non-Hermitian)  Riesz projection
\begin{equation}\label{eq:Qk}
Q_{\vec{k}} \ = \ \frac{1}{2 \pi \im} \int_{\cu{C}} \frac{1}{z - \wh{L}_{\vec{k}}} \di z
\end{equation}
is rank one at $\vec{k}=0$ and continuous as a function of $\vec{k}$.  It follows that $Q_{\vec{k}}$ is rank one as long as $\sigma(\wh{L}_{\vec{k}})$ does not intersect the contour $\cu{C}$.    Thus for small $\vec{k}$ the only spectrum of $\wh{L}_{\vec{k}}$ in a neighborhood of zero is a non-degenerate eigenvalue, with associated eigenvector in the one-dimensional range of $Q_{\vec{k}}$. 

Let us call the eigenvalue $E(\vec{k})$, and the associated normalized eigenvector $\Phi_{\vec{k}}$. Clearly $E(\vec 0)=0$ and $\Phi_{0}= \delta_{0}\otimes 1$. Since 
\begin{equation}
\wh{L}_{\vec{k}}Q_{\vec{k}} \ = \ E(\vec{k}) Q_{\vec{k}}
\end{equation}
we may compute the derivatives of $E(\vec{k})$ by differentiating $\wh{L}_{\vec{k}}$ and the projection.   
In particular, $\grad E(\vec{k})$ is given by the so-called Feynman-Hellman formula
\begin{equation}
\grad E(\vec{k}) \ = \ \ip{\Phi_{\vec{k}}, \grad \wh{L}_{\vec{k}} \Phi_{\vec{k}}},
\end{equation}
from which it follows that $\grad E(\vec 0) =0$ since $\grad \wh{L}_{\vec{k}} = \im \grad \wh{K}_{\vec{k}}$ is off-diagonal in the position basis on $\cu{H}_{0}$.  Similarly, we have
\begin{multline}
\partial_{i} \partial_{j} E(\vec{k}) \ = \ \ip{ \Phi_{\vec{k}}, \partial_{i}\partial_{j} \wh{L}_{\vec{k}} \Phi_{\vec{k}}}
 + \ip{\partial_{i} \wh{L}_{\vec{k}} \Phi_{\vec{k}}, 
(1- Q_{\vec{k}}) \wh{L}_{\vec{k}}^{-1} (1-Q_{\vec{k}}) \partial_{j} \wh{L}_{\vec{k}} \Phi_{\vec{k}}} \\+ \ip{\partial_{j} \wh{L}_{\vec{k}} \Phi_{\vec{k}}, 
(1- Q_{\vec{k}}) \wh{L}_{\vec{k}}^{-1} (1-Q_{\vec{k}}) \partial_{i} \wh{L}_{\vec{k}} \Phi_{\vec{k}}}
\end{multline}
The first term on the r.h.s.\ vanishes at $\vec{k}=0$ and the other two combine to give the identities claimed in the Lemma. Since $[\Gamma(0)]^{-1}$ is positive definite, it follows from the non-degeneracy condition (3b) that $\partial_{i} \partial_{j} E(\vec{0})$ is positive definite as well.
\end{proof}

Again, we obtain dynamical information about the semi-group $\e^{-t \wh{L}_{\vec{k}}}$:
\begin{lem}\label{lem:Lkdynamics} If $\epsilon$ is sufficiently small, then there is $C_{\epsilon } <\infty$ such that
\begin{equation}
\norm{\e^{-t \wh{L}_{\vec{k}}}(1- Q_{\vec{k}})} \le C_{\epsilon}\e^{-t (\delta_{\lambda} - \epsilon - c |\vec{k}|)}
\end{equation}
for all sufficiently small $\vec{k}$.
\end{lem}
\begin{proof}  Since $\num( \wh{L} _{\vec{k}}) \subset \cu{N}_{+}$, with $\cu{N}_{+}$ as in \eqref{eq:numrange} in the proof of Lemma \ref{lem:L0}, this 
is essentially identical to the proof of Lemma \ref{lem:L0dynamics}.  For $\vec{k}$ in a compact neighborhood of the origin, we can choose the bound $C_{\epsilon}$ uniform in $\vec{k}$.
\end{proof}

\subsection{Proof of Theorem \ref{thm:main}}
A first observation is that it suffices to prove diffusion \eqref{eq:main} under the assumption that the initial density matrix  $\rho_{0}$ satisfies
\begin{equation}\label{eq:summable}
\sum_{x,y} |\rho_{0}(x,y)| < \infty.
\end{equation}
To see this, it is useful to note:
\begin{enumerate}
\item The evolution \eqref{MAMDM} preserves the \emph{trace norm} of $\rho_{0}$.  
\item Any $\rho \in \cu{DM}$ may be approximated, to arbitrary precision, by an operator $\rho_{0}\in \cu{DM}$ satisfying \eqref{eq:summable}.
\end{enumerate}

In more detail, recall that the \emph{trace norm} of an operator $A$ is 
\begin{equation}
\norm{A}_{\cu{T}_{1}} \ = \ \sup \{ \abs{\tr A B} \ : \ B \text{ is finite rank and $\norm{B} \le 1$}\}.
\end{equation}
Since the evolution \eqref{MAMDM}  is given by $\rho_{t}  = U \rho_{0} U^{\dagger}$,
with $U = U(t,0)$ the unitary propagator of \eqref{MAM}, we see that $\norm{\rho_{t}}_{\cu{T}_{1}} = \norm{\rho_{0}}_{\cu{T}_{1}}$.  On the other hand, given $\rho_{0} $ of trace class and $\epsilon >0$ we can find $\wt{\rho}_{0} $ satisfying \eqref{eq:summable} and such that $\norm{\rho_{0}- \wt{\rho}_{0}}_{\cu{T}_{1}} < \epsilon$.  Indeed, we may take
\begin{equation}
\wt{\rho}_{0} (x,y) = \begin{cases} \rho_{0}(x,y) & \text{ if } |x|, |y| < L \\
0 & \text{ if } |x| \ge L \text{ or } |y| \ge L
\end{cases}
\end{equation}
with $L$ sufficiently large.  Then,
\begin{multline}
\abs{ \sum_{x} \e^{-\im \vec{k} \cdot x} \Ev{ \rho_{t}(x,x) - \wt{\rho}_{t}(x,x)}  }\\ \le \sum_{x} \Ev{ \abs{ \rho_{t}(x,x) - \wt{\rho}_{t}(x,x)}} \le \Ev{\norm{\rho_{t} - \wt{\rho}_{t}}_{\cu{T}_{1}}} \le \epsilon.
\end{multline}
If diffusion \eqref{eq:main} holds for any $\wt{\rho}_{0}$ satisfying \eqref{eq:summable} we learn that
\begin{equation}
\limsup_{\tau \ra \infty} 
\abs{ \sum_{x} \e^{-\im \frac{1}{\sqrt{\tau}}\vec{k} \cdot x} \Ev{ \rho_{\tau t}(x,x)} - \left ( \tr \rho_{0} \right ) \e^{- \sum_{i,j}D_{i,j}\vec{k}_{i}\vec{k}_{j} }} \le 2 \epsilon .
\end{equation}
which gives diffusion for $\rho_{0}$ in the limit $\epsilon \ra 0$. 

Turning now to $\rho_{0}$ which satisfies \eqref{eq:summable} we see that
\begin{equation}
\wh{\rho}_{0;\vec{k}}(x) \ = \ \sum_{y} \rho_{0}(x+y,y)\e^{\im \vec{k}\cdot y}
\end{equation}
defines a function that is uniformly bounded in $\ell^{2}(\Z^{d})$ as $\vec{k}$ varies through the torus,
\begin{equation}
\left [ \sum_{x}\abs{\rho_{0;\vec{k}}(x)}^{2} \right ]^{\half} \ \le \ \sum_{x}\abs{\rho_{0;\vec{k}}(x)} \ \le \  \sum_{x,y} \abs{\rho_{0}(x,y)} := M < \infty.
\end{equation}
By \eqref{E(Rho2)FT} we have
\begin{equation}
\sum_{x} \e^{-\im \frac{1}{\sqrt{\tau}} \vec{k} \cdot x} \Ev{\rho_{\tau t}(x,x)} \ = \  \ip{ \delta_{0} \otimes 1, \e^{-\tau t \wh{L}_{ \vec{k}/\sqrt{\tau}}} \wh{\rho}_{0;\frac{1}{\sqrt{\tau}}\vec{k}}\otimes 1}. \end{equation}
Letting $Q_{\vec{k}}$ denote the Riesz projection onto the eigenvector of $\wh{L}_{\vec{k}}$  near zero \tem \ see \eqref{eq:Qk} in the proof of Lemma \ref{lem:analytic} \tem, we have
\begin{multline}\label{eq:projectionsdecompose}
\sum_{x} \e^{-\im \frac{1}{\sqrt{\tau}} \vec{k} \cdot x} \Ev{\rho_{\tau t}(x,x)}  \ = \ \e^{-\tau t E(\vec{k}/\sqrt{\tau})} \ip{\delta_{0}\otimes 1, Q_{\frac{1}{\sqrt{\tau}}\vec{k}} \wh{\rho}_{0;\frac{1}{\sqrt{\tau}} \vec{k}} \otimes 1} \\ + \ \ip{ \delta_{0} \otimes 1, \e^{-\tau t \wh{L}_{ \vec{k}/\sqrt{\tau}}}  (1 - Q_{\frac{1}{\sqrt{\tau}}\vec{k}}) \wh{\rho}_{0;\frac{1}{\sqrt{\tau}}\vec{k}}\otimes 1},
\end{multline}
for $\tau$ sufficiently large.  By Lemma \ref{lem:Lkdynamics}, 
the second term in \eqref{eq:projectionsdecompose} is exponentially small in the large $\tau$ limit,
\begin{multline}\label{eq:decay}
\abs{\ip{ \delta_{0} \otimes 1,  (1 - Q_{\frac{1}{\sqrt{\tau}}\vec{k}}) \e^{-\tau t \wh{L}_{ \vec{k}/\sqrt{\tau}}} \wh{\rho}_{0;\frac{1}{\sqrt{\tau}}\vec{k}}\otimes 1}}
\\ \le  \  \norm{(1 - Q_{\frac{1}{\sqrt{\tau}}\vec{k}}) \e^{-\tau t \wh{L}_{ \vec{k}/\sqrt{\tau}}}}   \norm{  \wh{\rho}_{0;\frac{1}{\sqrt{\tau}}\vec{k}}\otimes 1} \
\le \ M  C_{\epsilon}  \e^{-\tau t (\delta_{\lambda}-\epsilon - c |\vec{k}|/\sqrt{\tau})} \ \ra \ 0.
\end{multline}

Regarding the first term in \eqref{eq:projectionsdecompose}, we have by Taylor's formula, 
\begin{equation}
E(\vec{k}/\sqrt{\tau}) \ = \ \frac{1}{2\tau} \sum_{i,j} \partial_{i} \partial_{j} E(\vec 0)  \vec{k}_{i} \vec{k}_{j} \ + \ o\left ( \frac{1}{\tau} \right ),
\end{equation}
since $E(\vec 0) = \nabla E(\vec 0)=0$. Thus
\begin{equation}\label{eq:taylor}
\e^{-\tau t E(\vec{k}/\sqrt{\tau})} \ = \ \e^{-t  \frac{1}{2} \sum_{i,j} \partial_{i} \partial_{j} E(\vec 0)  \vec{k}_{i} \vec{k}_{j}} + o(1).
\end{equation}
Putting together \eqref{eq:taylor} and \eqref{eq:decay} yields
\begin{multline}\label{eq:final}
\sum_{x} \e^{-\im \frac{1}{\sqrt{\tau}} \vec{k} \cdot x} \Ev{\rho_{\tau t}(x,x)}  \ = \ \e^{-t \half \sum_{i,j} \partial_{i} \partial_{j} E(\vec 0)  \vec{k}_{i} \vec{k}_{j}} \ip{\delta_{0}\otimes 1,  \wh{\rho}_{0;\frac{1}{\sqrt{\tau}} \vec{k}}\otimes 1} + o(1)  \\ = \ \e^{-t  \half \sum_{i,j} \partial_{i} \partial_{j} E(\vec0)  \vec{k}_{i} \vec{k}_{j}} \wh{\rho}_{0;\frac{1}{\sqrt{\tau}} \vec{k}}(0)+ o(1)
\ \ra \ \e^{-t  \half \sum_{i,j} \partial_{i} \partial_{j} E(\vec 0)  \vec{k}_{i} \vec{k}_{j}} \wh{\rho}_{0;\vec 0}(0)
\end{multline}
since $Q_{\vec{k}}^{\dagger} \delta_{0} \otimes 1 \ra \delta_{0} \otimes 1$ as $\vec{k} \ra 0$ and $\wh{\rho}_{0;\vec{k}}(0)$ is continuous as a function of $\vec{k}$.  

Eq.\ \eqref{eq:final} completes the proof of diffusion \eqref{eq:main}, with the diffusion matrix given by $D_{i,j}= \frac{1}{2} \partial_{i} \partial_{j} E(\vec 0)$.   
From the explicit expression \eqref{eq:explicit} for $\partial_{i} \partial_{j} E(\vec 0)$ in Lemma \ref{lem:analytic}, we see that the asymptotic form \eqref{eq:asdiff} holds, with
\begin{equation}
D_{i,j}^{0} =  \Re \Big \langle \partial_{i} \wh{K}_{\vec{0}} \delta_{0} \otimes 1, \left [ P_{0} \wh{V} P_{0}^{\perp} \left ( \im  \wh{K}_{\vec{0}} + B \right )^{-1} P_{0}^{\perp} \wh{V} P_{0} \right ]^{-1} \partial_{j} \wh{K}_{\vec{0}}\delta_{0} \otimes 1 \Big \rangle.
\end{equation}

To derive  \eqref{eq:x2}, diffusive scaling for $\sum_{x}\Ev{|x|^{2} \rho_{t}(x,x)}$, note that by \eqref{E(Rho2)FT} 
\begin{equation}
\frac{1}{t} \sum_{x} x^{2} \Ev{\rho_{t}(x,x)} \ = \ - \frac{1}{t} \Delta_{\vec{k}} \left . \big \langle \delta_{0} \otimes 1, \e^{-t \wh{L}_{\vec{k}}} \wh{\rho}_{0;\vec{k}} \otimes 1 \big \rangle \right |_{\vec{k}=0}.
\end{equation}
Expanding the r.h.s.\ we obtain
\begin{multline}\label{eq:x2expansion}
\frac{1}{t} \sum_{x} x^{2} \Ev{\rho_{t}(x,x)}  \ = \  \underbrace{- \frac{1}{t} 
\ip{\delta_{0} \otimes 1, \e^{-t \wh{L}_{\vec{0}}} \Delta_{\vec{k}} \wh{\rho}_{0;\vec{k}}|_{\vec{k}=0} \otimes 1 }}_{\mathrm{I}} \\
 \underbrace{- \frac{2}{t} \sum_{i=1}^{d}\ip{\delta_{0} \otimes 1,  \partial_{i}\left. \e^{-t \wh{L}_{\vec{k}}} \right  |_{\vec{k}=0} \partial_{i} \left . \wh{\rho}_{0;\vec{k}}  \right |_{\vec{k}=0}\otimes 1 }}_{\mathrm{II}}
\\ \underbrace{-\frac{1}{t} \ip{\delta_{0} \otimes 1,  \Delta_{\vec{k}}\left . \e^{-t \wh{L}_{\vec{k}}} \right  |_{\vec{k}=0}  \wh{\rho}_{0;\vec{0}} \otimes 1 }}_{\mathrm{III}}.
\end{multline}
The first term is negligible as $t \ra \infty$,
\begin{equation} 
\mathrm{I} \ = \ -\frac{1}{t} 
\ip{\delta_{0} \otimes 1, \e^{-t \wh{L}_{\vec{0}}} \Delta_{\vec{k}} \wh{\rho}_{0;\vec{k}} |_{\vec{k}=0} \otimes 1 }
\ = \ - \frac{1}{t} \sum_{x} x^{2} \rho_{0}(x,x)  \ = \ O \left ( \frac{1}{t} \right ),
\end{equation}
since $\e^{-t \wh{L}_{\vec{0}}^{\dagger}} \delta_{0} \otimes 1 = \delta_{0}\otimes 1$.

To proceed, we need to recall the formula for differentiating a semi-group,
\begin{equation}
\partial_{i} \e^{-t \wh{L}_{\vec{k}}}  \ = \  -\int_{0}^{t} \e^{-(t-s) \wh{L}_{\vec{k}}}\left (  \partial_{i} \wh{L}_{\vec{k}} \right ) \e^{-s \wh{L}_{\vec{k}}} \di s. 
\end{equation}
Thus the second term on the r.h.s.\ of \eqref{eq:x2expansion} is equal to
\begin{equation}
\mathrm{II} \ = \ - \frac{2}{t}  \int_{0}^{t } \sum_{i=1}^{d}\ip{\partial_{i} \wh{L}_{\vec{0}} \delta_{0} \otimes 1,  \e^{-s \wh{L}_{\vec{0}}}  (1-Q_{0}) \partial_{i}\wh{\rho}_{0;\vec{0}} \otimes 1 }
\di s,
\end{equation}
where we have recalled that
\begin{enumerate}
\item $\left [ \partial_{i}  \wh{L}_{\vec{k}} \right ]^{\dagger}= - \partial_{i}  \wh{L}_{\vec{k}}$, 
\item $\e^{-s \wh{L}_{\vec{0}}}Q_{0}=Q_{0}$, and
\item $Q_{0} \partial_{i}  \wh{L}_{\vec{0}} =0$.
\end{enumerate}
By Lemma \ref{lem:L0dynamics}, we see that $\mathrm{II}$ is negligible in the large $t$ limit:
\begin{equation}
\abs{\mathrm{II}} \ \lesssim \ \frac{1}{t}  \int_{0}^{\infty } \e^{-s (\delta_{\lambda} - \epsilon)} \di s \ = \ O \left (\frac{1}{t} \right ).
\end{equation}

It remains to compute the third term $\mathrm{III}$, for which we have
\begin{multline}
\mathrm{III} \ = \  
\underbrace{- \frac{1}{t} \int_{0}^{t}\ip{\Delta_{\vec{k}}  \wh{L}_{\vec{k}} |_{\vec{k}=0} \delta_{0} \otimes 1, \e^{-s \wh{L}_{\vec{0}}}  (1- Q_{0}) \wh{\rho}_{0;\vec{0}} \otimes 1 } \di s}_{\mathrm{IIIa}}
\\ + \underbrace{\frac{1}{t} \int_{0}^{t} \int_{0}^{s} \sum_{i=1}^{d}\ip{\partial_{i} \wh{L}_{\vec{0}} \delta_{0} \otimes 1, \e^{-(s-r) \wh{L}_{\vec{0}}} (1-Q_{0})\partial_{i} \wh{L}_{\vec{0}}  \e^{-r \wh{L}_{\vec{0}}} (1-Q_{0})\wh{\rho}_{0;\vec{0}} \otimes 1} \di r \di s}_{\mathrm{IIIb}} \\
+  \underbrace{\frac{1}{t} \int_{0}^{t} \int_{0}^{s} \sum_{i=1}^{d}\ip{\partial_{i} \wh{L}_{\vec{0}} \delta_{0} \otimes 1, \e^{-(s-r) \wh{L}_{\vec{0}}} (1-Q_{0})\partial_{i} \wh{L}_{\vec{0}}   Q_{0}\ \wh{\rho}_{0;\vec{0}} \otimes 1} \di r \di s}_{\mathrm{IIIc}}
\end{multline}
Similar to what we had for $\mathrm{II}$, we find that  $\mathrm{IIIa}=O\left ( \frac{1}{t} \right )$, and also that
\begin{equation}
\abs{\mathrm{IIIb}} \ \lesssim \ \frac{1}{t} \int_{0}^{\infty} \int_{0}^{s} \e^{-s(\delta_{\lambda}-\epsilon)} \di r \di s \ = \  O \left ( \frac{1}{t} \right ).
\end{equation}
Finally, for $\mathrm{IIIc}$ we have
\begin{multline}
\mathrm{IIIc} \ = \  \sum_{i=1}^{d}\ip{\partial_{i} \wh{L}_{\vec{0}} \delta_{0} \otimes 1, \wh{L}_{\vec{0}}^{-1} (1-Q_{0})\partial_{i} \wh{L}_{\vec{0}}   Q_{0}\ \wh{\rho}_{0;\vec{0}} \otimes 1} 
+ O\left ( \frac{1}{t} \right ), \\
= \ (\tr \rho_{0}) \sum_{i=1}^{d}\ip{\partial_{i} \wh{K}_{\vec{0}} \delta_{0} \otimes 1, \wh{L}_{\vec{0}}^{-1} \partial_{i} \wh{K}_{\vec{0}}   \delta_{0} \otimes 1} 
+ O\left ( \frac{1}{t} \right ),
\end{multline}
since $\nabla \wh{L}_{\vec{k}}= \im \nabla \wh{K}_{\vec{k}}$.  Comparing with the expression \eqref{eq:explicit} for $\partial_{i} \partial_{j}E(\vec{0})$ yields  \eqref{eq:x2}. \qed

\subsection*{Acknowledgments} We wish to express our gratitude for the hospitality extended to both of us by the Erwin Schr\"odinger Institute and to J.S. by the Isaac Newton Institute.  J. S. would like to thank Tom Spencer, J\"urg Fr\"ohlich and Wojhiech De Roeck for interesting discussions related to this work.  We would also like to thank Sergey Denissov for pointing out that the non-degeneracy condition (3b) on the hopping operator was missing from our original manuscript.

\providecommand{\bysame}{\leavevmode\hbox to3em{\hrulefill}\thinspace}
\providecommand{\MR}{\relax\ifhmode\unskip\space\fi MR }
\providecommand{\MRhref}[2]{%
  \href{http://www.ams.org/mathscinet-getitem?mr=#1}{#2}
}
\providecommand{\href}[2]{#2}

\end{document}